\documentstyle[12pt]{article}
\begin{document}
\renewcommand{\thefootnote}{\fnsymbol{footnote}}

\begin{center}
{\large\bf Friedmann Universes and Exact Solutions in String
Cosmology}\\
\bigskip

 S.N. Roshchupkin \\
 \medskip
 {\it Simferopol State University}\\
{\it 333036, Simferopol, Ukraina}\\  \vspace{5mm}

  and A.A. Zheltukhin\footnote{E-mail:kfti@kfti.kharkov.ua}\\ \vspace{5mm}

{\it Kharkov Institute of Physics and Technology}\\
{\it 310108, Kharkov, Ukraina}\\
\end{center}
\vspace{1.5cm}

\begin{quotation}
{\small\rm
We show that the classical null strings
 generate the Hilbert-Einstein gravity corresponding to D-dimensional
 Friedmann universes.
 } \end{quotation}

\renewcommand{\thefootnote}{\arabic{footnote}}
\setcounter{footnote}0
\vspace{1cm}

During last years much attention is devoted to the study of
different physical mechanisms of the inflation expansion in cosmological
backgrounds [1-5]. Recently the new string source of the
inflationary scenario has been considered [6-8] for the $D$-dimensional
Friedmann-Robertson-Walker (FRW) spaces describing by the metric
\begin{equation}\label{1}
ds^2=G_{MN}dx^Mdx^N=(dx^0)^2-R^2(x^0)dx^i\delta_{ik}dx^k ,
\end{equation}
where $~M, N= 0,1,...,D-1~$. Such a possibility is provided by
the existence of asymptotic string configuration describing an approximate
solution of the Hilbert-Einstein (HE) and string equations
in the large $~R(t)~$ limit. This solution corresponds to the phase of
unstable non-oscillating strings with proper sizes growing like the scale
factor$~R(t)~$ (1). In addition to the inflationary solution a new
approximate solution valid in the small $~R(t)~$ limit for a negatively
accelerated contraction $(d^2R/dt^2 < 0, ~ dR/dt < 0)~$ of the universes
was found in [8]. The last solution describes the phase of perfect gas of
shrunk strings $~(R(t)\rightarrow 0,~ \tau \to0)~$ with the approximate
equation of state
\begin{equation}\label{2} \rho \approx p(D-1). \end{equation}

Here we want to show that the classical equations for null strings
together with gravity equations are exactly solvable for the case of the
FRW metric (1). These solutions correspond to the phase of perfect gas of
shrunk or stretched $~(dR/dt>0, d^2R/dt^2<0)~$
null strings with the exact equation of state
\begin{equation}\label{3}
\rho = p(D-1), \end{equation} Due to this fact we may think that the
perfect    gas of null strings is an alternative source of the FRW gravity
with $~k=0.$ To prove this result note that the action for null string in
a cosmological background $~G_{MN}(x)~$ may be written as [9,10]
\begin{equation}\label{4} S=\int d\tau d\sigma {det(\partial_{\mu} x^M
G_{MN}(x) \partial_{\nu} x^N) \over E(\tau,\sigma )}, \end{equation} In
the world-sheet gauge fixed by the conditions [9,10]
\begin{equation}\label{5} \dot{x}^M G_{MN} x^{\prime N} =0, ~~~~~~~ {
x^{\prime M} G_{MN}(x) x^{\prime N} \over E(\tau\sigma)} = {1 \over
\gamma^\ast \pi}~, \end{equation} where $~\gamma ^\ast~$--
constant with the dimension $L^2~$ (if $~\hbar = c =1~$), the motion
equations and constraints produced by $~S~$ (4) have the following form $$
\ddot{x}^M + \Gamma_{PQ}^M \dot{x}^P \dot{x}^Q = 0 $$
\begin{equation}\label{6} \dot{x}^M G_{MN}\dot{x}^N =0,~~~~~~~~ \dot{x}^M
G_{MN} x^{\prime N} =0, \end{equation} where $~\dot{x}=\partial x
/\partial \tau, ~~ x^\prime = \partial x /\partial \sigma.~~$ To solve
Eqs. (6) it is convenient to turn to a conformal time $~\tilde{x}^0(\tau
,\sigma )~$, defined by \begin{equation}\label{7}
dx^0=C(\tilde{x}^0)d\tilde{x}^0,~~~~~C(\tilde{x}^0) = R(x^0),~~~~~
\tilde{x}^i = x^i \end{equation} In the gauge of the conformal time the
metric (1) is presented in the conformal-flat form
\begin{equation}\label{8} ds^2=C(\tilde{x}^0) \eta_{MN}d\tilde{x}^M
d\tilde{x}^N,~~~~~~~~~ \eta_{MN}= diag(1, -\delta_{ij}) \end{equation}
with the Christoffel symbols
$~\tilde{\Gamma}_{PQ}^M(\tilde{x})~~$ [11] \begin{equation}\label{9}
\tilde{\Gamma}_{PQ}^M(\tilde{x})= C^{-1}(\tilde{x})[\delta_P^M
\tilde{\partial}_QC+\delta_Q^M \tilde{\partial}_PC -
\eta_{PQ}\tilde{\partial}^MC] \end{equation} Taking into account the
 relation (8,9) we transform Eqs. (6) to the form
 \begin{equation}\label{10} \ddot{\tilde{x}}^M +
 2C^{-1}\dot{C}\dot{\tilde{x}}^M=0 \end{equation}
 \begin{equation}\label{11}
 \eta_{MN}\dot{\tilde{x}}^M\dot{\tilde{x}}^N=0,~~~~~~~~\eta_{MN}
 \dot{\tilde{x}}^M\tilde{x}^{\prime N}=0
 \end{equation}
 The first integration of Eqs. (10) results in the first order equations
 \begin{equation}\label{12}
 H^\ast C^2\dot{\tilde{x}}^0 = \psi^0(\sigma ),~~~~~~~ H^\ast C^2
 \dot{\tilde{x}}^i= \psi^i(\sigma ),
 \end{equation}
 the solutions of which have the form
\begin{equation}\label{13}
\tau = H^\ast \psi_0^{-1} \int_{t_o}^t
 dt R(t), \end{equation}
 $$ x^i(\tau ,\sigma ) = x^i(0,\sigma ) + H^{\ast
 -1} \psi^i \int_0^\tau d\tau R^{-2}(t),$$
where $~H^{\ast}~$ is a metric constant with the dimension $~L^{-1}~$ and
$~t_0 \equiv x^0(0,\sigma ), ~~x^i(0,\sigma )~$ and $~ \psi ^M(\sigma
 )~$ are the initial data. The solution (13) for the space world
 coordinates $~x^i(t) (i=1,..., D-1)~$ as a function of the cosmic time $
 ~t=x^0~$, may be written in the equivalent form as
 \begin{equation}\label{14}
 x^i(t,\sigma )=x^i(t_0,\sigma )+\nu^i(\sigma )\int_{t_o}^t~ dt R^{-1}(t),
 \end{equation}
 where
 $~\nu^i(\sigma ) \equiv \psi_o^{-1}\psi^i.$

 The explicit form of the solutions (13) allows to transform the
 constraints (11) into those for the Caushy initial
 data: $~ \nu^i(\sigma ), ~ t_0(\sigma)~$ and
 $~x^{\prime}_0 \equiv x^{\prime}(t_0,\sigma )$ \begin{equation}\label{15}
 \nu^i(\sigma )\nu^k(\sigma )\delta_{ik} =1,~~ t_0^\prime (\sigma ) =
R(t_0)(x_0^{\prime i} \nu^k\delta_{ik}). \end{equation} The
energy-momentum tensor $~ T^{MN}(x)~$ of null string is defined by the
variation  of the action (4) with respect  to $~ G_{MN}(x)~$
 \begin{equation}\label{16} T^{MN}(X)= {1 \over
\pi\gamma^\ast \sqrt{ _{|G|}}} \int ~d\tau d\sigma
\dot{x}^M\dot{x}^N\delta^D(X-x).\end{equation} The non-zero $~T_{MN}~$
components (16) have the following form
$$
T^{00}(X)=\frac{1}{\pi\gamma^*
H^*}R^{(-D)}(t)\int d\sigma
\psi_0(\sigma)\delta^{D-1}(X^M-x^M(\tau,\sigma)),
$$
\begin{equation}\label{17}
T^{ik}(X)=\frac{1}{\pi\gamma^*H^*}R^{(-D-2)}\int
d\sigma\nu^i(\sigma)\nu^k(\sigma)\psi_0(\sigma)\delta^{D-1}
(X^M-x^M(\tau,\sigma),
\end{equation} where the time dependence $~T^{MN}~$ is factorized and
accumulated in the scale factor $~R(t).~$ Since the vector $~\dot{x}^M~$
is a light-like one, the trace $~T^{MN}~$ equals to zero
\begin{equation}\label{18}
Sp~T ={T_0}^0+G_{ij}T^{ij}=0 . \end{equation}

To consider the null strings (17) as a source of the FRW gravity (1) it is
convenient to pass from the separate null string to the perfect gas of
these strings, supposing that this gas is homogenious and isotropic. Then
the energy density $~\rho (t)~$ and pressure $~p(t)~$ of the gas and its
energy-momentum $~\langle T^{MN}(x)\rangle~$ are connected by the
relations
\begin {equation}\label{19} \langle {T_0} ^0\rangle =\rho(t),~~~~~~~~
\langle {T_i} ^j\rangle   = -p(t){\delta_i} ^j.\end{equation}
The tensor $~\langle T_{MN}\rangle~$ is derived from  $~T_{MN}~$ by means of
its
space averaging when a set of null strings is introduced instead of a
single null string. As a result of this procedure we find that $~\rho~$
and $~p~$ are
\begin {equation}\label{20}
\langle {T_0 }^0 \rangle = \rho(t) =  {A \over R^D(t)},~~~~~~
\langle {T_i }^j \rangle = -p(t){\delta_i} ^j = - {
{\delta_i}^j \over D-1} ~{A\over R^D(t)}, \end{equation} where $~A~$ is a
constant with the dimension $~L^{-D}~$.  Eqs.(20) show that the equation
of state of null string gas is just the equation of state for a gas of
massless particles \begin {equation}\label{21} \langle Sp~T \rangle =
\langle {T_M} ^M \rangle =0 \Longleftrightarrow \rho =
(D-1)p.\end{equation} Now assume that the gas of null strings is a
dominated source of the FRW gravity (1).  For the validity of the last
conjecture it is necessary that the HE equations \begin
{equation}\label{22} {R_M} ^N = 8\pi G_D\langle {T_M}^N \rangle
   \end{equation} with the non-zero Ricci tensor $~{R_M} ^N~$ components
defined by $~{G_M} ^N~$ (1) $$~ {R_0} ^0 = -{D-1 \over R}~{d^2R\over
dt^2}, $$ \begin{equation}\label{23} {R_i} ^k = -{\delta _i} ^k
\left[{1\over R}~ {d^2R\over dt^2} + {D-2\over R^2} \left( {dR\over
 dt}\right)^2\right] \end{equation} should contain the tensor
$\langle{T_M} ^N\rangle$ (20) as a source of the FRW gravity.  Moreover,
the constraints (20), i.e. $$ \rho R^D - A = 0 $$ must be a motion
integral for the HE system (22). It is actually realized because $$
{d\over dt}(\rho R^D) = - {D-2\over 16 \pi G_D } R^{D-1}~{dR\over dt }
{R_M} ^M = 0, $$ since the trace  $~{R_M}^M \sim \langle{T_M}^M\rangle=0~
$ (see (21)). In view of this fact it is enough to consider only one
equation of the system (22) \begin{equation}\label{24} \left({1\over
R}~{dR \over dt}\right)^2 = {16 \pi G_D \over(D-1)(D-2)} ~{A\over R^D}\ \
, \end{equation} which defines the scale factor R(t) of the FRW metric
(1). Note that in the case  $D~=~4$ Eq. (24) turns into the well-known
Friedmann equation for the energy density in the radiation dominated
universe with  $~k = 0~$.  The solutions of Eq. (24) are $$ R_I(t) =
[q(t_c -t)]^{2/D}, ~~~~~~~ t < t_c, $$ $$ R_{II}(t) =[q(t
-t_c)]^{2/D},~~~~~~ t > t_c , $$ where $q=[4\pi G_D A /
(D-1)(D-2)]^{1/2}~~$ and $t_c$ is a constant of integration which is a
singular point of the metric (1). The solution $~R_I~$ describes the stage
of negatively accelerated contraction of D-dimensional FRW universe. In
the small $R$ limit $(R\rightarrow 0)$ this solution was found in [8] as
an approximate asymptotic solution for the gas of strings with non-zero
tension. For the case of null strings this solution is exact. The second
solution (24) $~R_{II}~$ describes the stage of the negatively accelerated
expansion of the FRW universe from the state with space volume equals to
"zero".  Thus we see that the perfect gas of noninteracting null strings
     may be considered as an alternative source of the gravity in the FRW
universes with $~k = 0~$.  From the view point of string cosmology it
seems important to find other exclusive metrics selfconsistently connected
with the dynamics of null strings and allowing  to consider the latter as
a dominant source of gravity. The most interesting are the metrics which
could describe the inflation of spaces. The solution of the problem under
question is divided into two steps. At the first step it is necessary to
exactly solve the null string motion equation in some  cosmological
background. The existence of such solution produces the constraints
between $~T_{MN}~$  and $~G_{MN}~$.  From this moment the source in the HE
equations becomes fixed as function on the background metric $~G_{MN}~$.
At the second step we must solve these HE equations.  The number of the
independent functions characterizing $~G_{MN}~$ may be less than the
number of the independent HE equations. Then the condition of  the
selfconsistency of the considered scenario demands that the constraints
(or some part of them) between $~T_{MN}~$ and $~G_{MN}~$ should be
conserved integrals of the HE equations.  The last condition may turn out
to be incorrect. Then we must deform our initial background metric and
again repeat the first and second steps with the new metric
$\tilde{G}_{MN}$.  As an example of above discussed inconsistency let us
consider the $~(d+1)~$-dimensional FRW space $~(t,x^i)~$ extended by the
addition of $~n~$ internal compactified dimensions $~y^a~$ [8].
Choose the metric $~G_{MN}(z^L)~$ (where $~z^L=(t,x^i,y^a)~$) of this
extended space--time in the form [8]:
\begin{equation}\label{25}
ds^2=G_{MN}dx^Mdx^N=(dt)^2-R^2(t)\sum_{ik=1}^{d}dx^i\delta_{ik}dx^k -
r^2(t)\sum_{a,b=d+1}^{d+n}dy^a\delta_{ab}dy^b .
\end{equation} In the metric (25) the classical equation of motion (6)
(where $~z^M~$ is substituted instead of $~x^M~$) are exactly integrable,
so that $$
\tau=\int_{t_0}^t\frac{dt}{\sqrt{\lambda(\sigma)+AR^{-2}+ar^{-2}}},
$$
$$
x^i(\tau,\sigma)=x^i(0,\sigma)+\mu^i(\sigma)\int_0^{\tau}d\tau R^{-2}(t),
$$
\begin{equation}\label{26}
y^a(\tau,\sigma)=y^a(0,\sigma)+\mu^a(\sigma)\int_0^{\tau}d\tau r^{-2}(t),
\end{equation}
where $~\lambda~$,$~\mu^i~$,$~\mu^a~$ and
$~A=\mu^i\delta_{ik}\mu^k~$,$~a=\mu^a\delta_{ab}\mu^b~$ are the
"constants" of integration and their functions. The substitution of the
solution (26) into the constraints (6) transforms them into the
constraints for the initial data
\begin{equation}\label{27}
t^{\prime}(\sigma)=\frac{\mu^i\delta_{ik}{x^{\prime}}_0 ^k +
\mu^a\delta_{ab}{y^{\prime}}_0^a}{\sqrt{AR^{-2}(t_0)+ar^{-2}(t_0)}},~~~~~~
\lambda(\sigma)=0
\end{equation}

The non--zero components of the energy--momentum tensor $~{T_M}^N~$ (16)
corresponding to the solutions (26) have the form
$$
{T_0}^0(Z)=\frac{1}{\pi\gamma^*}\int d\tau d\sigma
R^{-d}r^{-n}(AR^{-2}+ar^{-2})\delta^D(Z^M-z^M(\tau,\sigma)),
$$
$$
{T_i}^k(Z)=-\frac{1}{\pi\gamma^*}\int d\tau d\sigma
R^{-(d+2)}r^{-n}\mu_i\mu^k\delta^D(Z^M-z^M(\tau,\sigma)),
$$
\begin{equation}\label{28}
{T_a}^b(Z)=-\frac{1}{\pi\gamma^*}\int d\tau d\sigma
R^{-d}r^{-(n+2)}\mu_a\mu^b\delta^D(Z^M-z^M(\tau,\sigma)) .
\end{equation}  As in the case of the FRW metric (1), the components of
$~{T_M}^N~$ are
connected by the condition $~{T_M}^M=0~$. The time dependence of
$~{T_M}^N~$ (28) is factorized and becomes evident after the integration
with respect to $~t~$ using the relation \\
$~dt=\sqrt{AR^{-2}+ar^{-2}}d\tau~$. As in the previous case, the
transition to the perfect gas of non--interacting null strings is realized
by the space averaging $~{T_M}^N~$ (28). As a result of this averaging we
obtain the following expressions for the energy density $\rho$ and
pressures $p$ and $q$ acting in $x$-- and $y$--subspaces
$$
<{T_0}^0>=\rho=\frac{1}{\gamma R^dr^n}\sqrt{AR^{-2}+ar^{-2}},
$$
$$
<{T_i}^k>=-p\delta_i^k=-\frac{\delta_i^k}{\gamma R^dr^n} \frac{A}{R^2d}
\frac{1}{\sqrt{AR^{-2}+ar^{-2}}},
$$
\begin{equation}\label{29}
<{T_a}^b>=-p\delta_a^b=-\frac{\delta_a^b}{\gamma R^dr^n} \frac{a}{r^2n}
\frac{1}{\sqrt{AR^{-2}+ar^{-2}}}.
\end{equation}
Then we find the equation of state
\begin{equation}\label{30}
\rho=dp+nq,
\end{equation}
characterizing the perfect gas of null strings in the metric (25). The HE
equations for the metric (25) have the form
$$
\frac{d}{R}\frac{d^2R}{dt^2}+ \frac{n}{r}\frac{d^2r}{dt^2}= -8\pi G_D\rho,
$$
$$
\frac{1}{R}\frac{d^2R}{dt^2}+
\frac{d-1}{R^2}\left(\frac{dR}{dt}\right)^2+\frac{n}{rR}\frac{dR}{dt}
\frac{dr}{dt}= 8\pi G_Dp,
$$
\begin{equation}\label{31}
\frac{1}{r}\frac{d^2r}{dt^2}+
\frac{n-1}{r^2}\left(\frac{dr}{dt}\right)^2+\frac{d}{rR}\frac{dR}{dt}
\frac{dr}{dt}= 8\pi G_Dq,
\end{equation}
where the expressions (29) for $\rho$, $p$ and $q$ are implied. The system
(31) consists of three equations for two functions $R(t)$ and $r(t)$.
Unlike the Friedmann system (22), the constraint (29) between $\rho$ and
$R$ is non conserved by the dynamics defined by Eqs.(31). Therefore the
perfect gas of mull strings can not be considered as a dominant gravity
source for the space--time (25).

Returning to the case of the FRW universes note that null strings may
apparently be considered as an ordered soliton--like state of a large
number of coherent photons. The formation of such string--like condensate
of radiation can reflect the possible spontaneous breaking of the
space--time symmetry in the FRW universes.

The second conclusion concerns the possibility of the generalization of
null string cosmology to the null $p$--brane [9,10] cosmology.  Such a
generalization is actually possible because the classical dynamics of null
$p$--branes is described by the similar (6) equations of motion together
with trivially extended system of constraints. Therefore the results
derived here for the gas of null strings in the FRW universes are
conserved also and for the perfect gas of null $p$--branes in the FRW
spaces.

A new moment connected with the null $p$--branes is that they can be
considered as $p$--dimensional ordered structures, in contrast to the
1--dimensional null string structures. So the world hypervolumes of the
null $p$--branes may be discussed as $(p+1)$--dimensional galaxies. When
null $p$--brane achieves the horizon it is splitted into a large number of
"small" pieces which are detected as the separate minihalaxies with the
practically identical physical characteristics. May be the consideration
of the visible galaxies as such pieces of null $p$--branes could shed new
light in solution of the horizon problem.

The authors wish to thank D.V.Volkov, Yu.P.Stepanovskij,
Yu.P.Peresun'ko,\\ A.P.Rekalo and B.I.Savchenko for the interest to this
work and useful discussions.

This work was supported in part by the International Science Foundation
(grant RY9200), International Association for the Promotion of Cooperation
with Scientists from the independent states of the former Soviet Union
(grants INTAS-93-127, 93-633, 94-2317) and by the Ukrainian State
Committee in Science and Technologies, Grant N 2.3/664.

\newpage
\section*{References}
\begin{enumerate}
\item A.D.Linde, Physics of elementary particles and inflation cosmology,
 \\ Nauka, Moscow, 1990.
\item K.A.Olive, Phys.Rep.{\bf 190}
 307 (1990).  \item G.Veneziano, Strings, cosmology, ... and a
 particle, CERN Preprint TH.7502/94.
\item N.Sanchez and G.Veneziano,
 Nucl.Phys.{\bf B333} 253 (1990).
\item M.Gasperini, N.Sanchez and
 G.Veneziano, Highly unstable fundamental strings in inflationary
 cosmologies, CERN preprint TH.5893/90.
\item M.Gasperini and G.Veneziano, Mod.Phys.Lett. {\bf A8} 3701 (1993)\\
Phys.Rev. {\bf D50} 2519 (1994).
\item M.Gasperini, N.Sanchez and
 G.Veneziano, Self-sustained inflation and dimensional reduction from
 fundamental strings, CERN preprint TH.5010/91
\item A.A.Zheltukhin, Pis'ma Zhur.Eksp.Teor.Fiz.
 {\bf 46} 262 (1987) (in Russian); Yad.Fiz. {\bf 48} 375 (1988) (in Russian).
\item I.A.Bandos and A.A.Zheltukhin, Fortschr.  Phys.{\bf41} 619 (1993).
\item A.Z.Petrov, New methods in general relativity, Nauka, Moscow, 1966.
 \end{enumerate}

\end{document}